\documentclass[11pt]{article}
\usepackage[english]{babel}
\usepackage{psfrag}
\usepackage{graphicx,subfigure} 
\usepackage{float,times}
\usepackage{vmargin}

\setpapersize{A4}
\setmarginsrb{1.0cm}{1.0cm}{1.0cm}{1.0cm}%
             {0em}{0em}{0em}{0cm}
\begin{document}

\twocolumn[\hsize\textwidth\columnwidth\hsize\csname
@twocolumnfalse\endcsname {\centerline{\large{\bf {Analysis of data
        sets of stochastic systems}}}} \vspace{0.3cm}
{\centerline{S.
    Siegert$^{a}$\footnotemark[1],
    R. Friedrich$^a$, J. Peinke$^{b}$\footnotemark[2] }} {\footnotesize
  \centerline{{\it $ ^a$ Institut f\"ur Theoretische Physik,
      Universit\"at Stuttgart, 70550 Stuttgart, Germany}}
  \centerline{{\it $ ^b$ Experimentalphysik II, Universit\"at
      Bayreuth, 95447 Bayreuth, Germany}}
  \centerline{(Submitted to Physics Letters,
    March 19, 1998)} } \vspace{0.2cm} {\footnotesize This paper deals
  with the analysis of data sets of stochastic systems which can be
  described by a Langevin equation. By the method presented in this
  paper drift and diffusion terms of the corresponding Fokker-Planck
  equation can be extracted from the noisy data sets, and
  deterministic laws and fluctuating forces of the dynamics can be
  identified. The method is validated by the application to simulated
  one- and two-dimensional noisy data
  sets. \\
  PACS number(s): 02.50.Fz, 02.50.Ey, 02.50.Ga }
\\
\\
]

\vspace{1cm}

\makeatletter
\footnotetext[1]{silke@theo3.physik.uni-stuttgart.de}
\footnotetext[2]{joachim.peinke@uni-bayreuth.de}
\makeatother

{\centerline {\bf I. INTRODUCTION}}
\vspace{0.3cm}

In biological, economical, physical or technical systems noisy data
sets occur very frequently. For describing and/or influencing these
complex systems, it is necessary to know the deterministic laws and
the strength of the fluctuations controlling the dynamics. Especially
in the dynamics of order parameters close to instability points
fluctuations play an important role \cite{Haken1}, \cite{Haken2},
\cite{Haken3}. This paper discusses a ge\-neral method which allows
one to determine drift and diffusion coefficients of the Fokker-Planck
equation for stationary continuous complex systems with Markov
properties. So a complete description of the stochastic process is
found. This problem has also been addressed by \cite{Borland1} and
\cite{Klimo1}, but in a different way.

\vspace{1cm}

{\centerline {\bf II. BASIC RELATIONS}}
\vspace{0.3cm}

{\centerline {\bf A. Langevin equation and Markovian property}}

The dynamics of a continuous Markovian system is governed by a
Langevin equation (\ref{Lan}) for a set of n variables $\{q_k(t)\}$,
$k=1, \dots n$ \cite{Kampen}:

\begin{eqnarray}\label{Lan}
  \frac{d}{dt}q_i(t)&=&h_i(\{q_k(t)\},t)+\sum_j g_{ij}(\{q_k(t)\},t)
  \Gamma _j(t) \nonumber \\
  && \\
  k&=&1, \dots n \nonumber
\end{eqnarray}

The fluctuating Langevin forces $\Gamma _j(t)$ are considered to be
$\delta$-correlated noise functions with vanishing mean as expressed
in equation (\ref{Korrel1}) and (\ref{Korrel2}).

\begin{eqnarray}\label{Korrel1}
  <\Gamma_i(t)> \quad &=& \makebox[2.5cm][c]{$0$} \,\, \forall \, i \\
    && \nonumber \\
    \label{Korrel2}
    <\Gamma_i(t) \Gamma_j(t')> &=& \makebox[2.5cm][c]{
      $Q \cdot \delta_{ij}\,\delta(t-t')$} \,\, \forall \, i,j
\end{eqnarray}

The condition of Markovian property demands that the dynamics of the
process depends only on the present state of the system and not on its
history.  With definition (\ref{Cond}) for conditional probability
density distributions $p$, the condition for the validity of Markovian
property can be expressed by equation (\ref{Markov}) \cite{Honerkamp},
\cite{Risken}.

\begin{eqnarray}\label{Cond}
  p(q_n,t_n &\mid& q_{n-1},t_{n-1};\dots;q_1,t_1) \equiv \nonumber \\
  && \nonumber \\
  &&\frac{w(q_n,t_n;\dots;q_1,t_1)}
  {w(q_{n-1},t_{n-1};\dots;q_1,t_1)}
\end{eqnarray}

\noindent
$\int w(q_n,t_n;\dots;q_1,t_1)\prod_i dq_i$ is the probability to find
the system in $q_i \dots q_i+dq_i$ at time $t_i$ for all $i$.

\begin{equation}\label{Markov}
  p(q_n,t_n \mid q_{n-1},t_{n-1};\dots;q_1,t_1)
  = p(q_n,t_n \mid q_{n-1},t_{n-1}),
\end{equation}

\vspace{-0.2cm}
\noindent
\centerline{$t_n>t_{n-1}> \dots >t_1,$}

\noindent
where $t_{n-1}$ is the next earlier time before $t_n$.

This condition can be validated by analysing the data sets
numerically, as has been shown by \cite{Friedrich2}. In case
that the Markovian property does not fit, the dimension of the state
vector can be increased by including a further observable. This can
always be achieved by introducing delay coordinates. Thus by
introducing new random variables, non-Markovian processes may be
reduced to Markovian systems \cite{Risken}.

\vspace{0.3cm}

{\centerline {\bf B. Fokker-Planck equation}}

Usually a formal general solution of the stochastic differential
Langevin equation (\ref{Lan}) cannot be given. Therefore a
Fokker-Planck equation (\ref{Fokker}) is set up by which the
probabi\-lity density distribution $w(\{q_k(t),t\})$ of the stochastic
variables can be calculated \cite{Schuss}:

\begin{eqnarray}\label{Fokker}
  \lefteqn{
    \frac{\partial}{\partial t}w(\{q_k(t),t\})} \nonumber \\
  &=&
  -\sum^n_{i=1} \frac{\partial}{\partial q_i}
  \left( D_i^{(1)}(\{q_k(t)\},t)w(\{q_k(t)\},t) \right) \nonumber \\
  &+&\frac{1}{2} \sum_{ij=1}^n
  \frac{\partial ^2}{\partial q_i \partial q_j}
  \left( D_{ij}^{(2)}(\{q_k(t)\},t)w(\{q_k(t)\},t)\right)
\end{eqnarray}

\begin{displaymath}
  k=1,\dots n \hspace{5cm}
\end{displaymath}

The coefficients $D_i^{(1)}$ are called drift coefficients, the terms
$D_{ij}^{(2)}$ diffusion coefficients, they are defined in equation
(\ref{Drift1}) and (\ref{Diff1}) \cite{Risken}.

\begin{eqnarray}\label{Drift1}
  D_i^{(1)}(\{q(t)\},t)&=& 
  \lim_{\tau \to 0}\frac{1}{\tau}\langle \tilde q_i(t+\tau) -
  q_i \rangle |_{\{\tilde q_k(t)\}=\{q\}} \nonumber \\
  &&\nonumber\\
  &=& \lim_{\tau \to 0}\frac{1}{\tau}\int\limits_{-\infty}^{+\infty}
  \left(\tilde q_i(t+\tau)- q_i \right)\cdot \nonumber \\
  && \nonumber \\
  && p(\{\tilde q_k\},t+\tau \mid \{q_k\},t)
  \prod_k d\tilde q_k \\
  && \nonumber \\
  && \nonumber \\
  \label{Diff1}
  D_{ij}^{(2)}(\{q_k(t)\},t)&=&
  \lim_{\tau \to 0}\frac{1}{\tau}\langle (\tilde q_i(t+\tau) - q_i)
  \cdot \nonumber \\
  && \nonumber \\
  && (\tilde q_j(t+\tau) - q_j)\rangle
  |_{\tilde \{q_k(t)\}=\{q_k\}} \nonumber \\
  && \nonumber \\
  &=& \lim_{\tau \to 0}\frac{1}{\tau}\int\limits_{-\infty}^{+\infty}
  \left(\tilde q_i(t+\tau)-q_i\right) \cdot \nonumber \\
  &&\left(\tilde q_j(t+\tau)-q_j\right) \cdot \nonumber \\
  && \nonumber \\
  && p(\{\tilde q_k\},t+\tau \mid \{q_k\},t) \prod_k d\tilde q_k,\\
  k&=&1,\dots n \nonumber
\end{eqnarray}

\noindent
where $\{\tilde q_k(t+\tau)\}$ with $\tau > 0$ is a solution of
(\ref{Lan}) which at time $t$ has the sharp value $\{ \tilde
  q_k(t)\}=\{q_k\}$.

In \cite{Friedrich1} an application of these equations (\ref{Drift1})
and (\ref{Diff1}) together with the definitions of conditional
probability density distributions (\ref{Cond}) to the statistical
properties of a turbulent cascade is discussed.

The relations between the sets of coefficients of the Lan\-ge\-vin
equation and the Fokker-Planck equation in consideration of the
Stratono\-vich definitions are \cite{Risken}

\begin{eqnarray}\label{Drift2}
  \lefteqn{
  D_i^{(1)}(\{q_k(t)\},t)=h_i(\{q_k(t)\},t)} \nonumber \\
  && \nonumber\\
  &&+\frac{Q}{2}
  \sum_{l,j} g_{lj}(\{q_k(t)\},t)\frac{\partial}{\partial q_l}
  g_{ij}(\{q_k(t)\},t) \nonumber\\
  && \\
  &&\nonumber \\
  \label{Diff2}
  \lefteqn{
    D_{ij}^{(2)}(\{q_k(t)\},t)} \nonumber \\
  && \nonumber \\
  && =Q\cdot \sum_l g_{il}(\{q_k(t)\},t)
  g_{jl}(\{q_k(t)\},t)\\
  && \nonumber \\
  &&k=1,\dots n \nonumber
\end{eqnarray}

\vspace{1cm}

\newpage
{\centerline {\bf III. ANALYSIS OF STOCHASTIC DATA SETS}}
\vspace{0.3cm}

A continuous Markovian system in the presence of white noise is
governed by a Langevin equation (\ref{Lan}) respectively by the
corresponding Fokker-Planck equation (\ref{Fokker}). A complete
analysis of such a complex system, therefore, should yield the
quantities $h_i$ and $g_{ij}$ of the Langevin equation (\ref{Lan})
respectively drift and diffusion coefficients $D_i^{(1)}$ and
$D_{ij}^{(2)}$ of the Fokker-Planck equation (\ref{Fokker}). Both sets
of terms are according to (\ref{Drift2}) and (\ref{Diff2}) correlated
with each other.

For the regarded class of stationary continuous
Markovian systems with white dynamical noise, where the validity of
the Markovian property may have been achieved by the introduction of
delay coordinates, it is always possible to determine drift and
diffusion terms directly from the data sets by using the equations
(\ref{Cond}), (\ref{Drift1}) and (\ref{Diff1}). Because of the
stationarity, $D_i^{(1)}$ and $D_{ij}^{(2)}$ have no explicit time
dependence. The needed conditional probability density distribution
can be determined numerically from the data set according to
(\ref{Cond}) by calculating histograms.

This general algorithm has apparently not been recognized in
literature up to now. The numerical method is completely general, no
ansatz for the coefficients is needed. If analytical formulas for the
evolution equations of a process are needed, the gained numerical
results may be approximated by analytical functions.

\newpage
%\vspace{1cm}

{\centerline {\bf IV. APPLICATIONS OF THE PRESENTED}}
{\centerline {\bf ALGORITHM}}
\vspace{0.3cm}

In the following the discussed algorithm will be applied to various
one- and two-di\-men\-sio\-nal syn\-the\-ti\-cally de\-ter-\linebreak
mined data sets.

\vspace {0.3cm}

{\centerline {\bf A. First example}}
\vspace{0.3cm}

\begin{figure}[H]
  \begin{center}
    {\begin{psfrags}
% Achsenbeschriftung
        \psfrag{x}[cc][]{{\footnotesize Variable q(t)}}
        \psfrag{t}[cc][]{{\footnotesize Time t}}
% y-Achse
        \psfrag{0.2}[r][]{{\scriptsize 0.2}}
        \psfrag{0.4}[r][]{{\scriptsize 0.4}}
        \psfrag{-0.2}[r][]{{\scriptsize -0.2}}
        \psfrag{-0.4}[r][]{{\scriptsize -0.4}}
        \psfrag{0y}[r][]{{\scriptsize 0\,}}        
% x-Achse
        \psfrag{0}[t][]{{\scriptsize 0}}        
        \psfrag{1000}[t][]{{\scriptsize 1000}}
        \psfrag{2000}[t][]{{\scriptsize 2000}}
        \psfrag{3000}[t][]{{\scriptsize 3000}}
        \psfrag{4000}[t][]{{\scriptsize 4000}}
        \psfrag{5000}[t][]{{\scriptsize 5000}}
% Groesse des Bildes
        \includegraphics[width=5cm,height=8.0cm,angle=270]{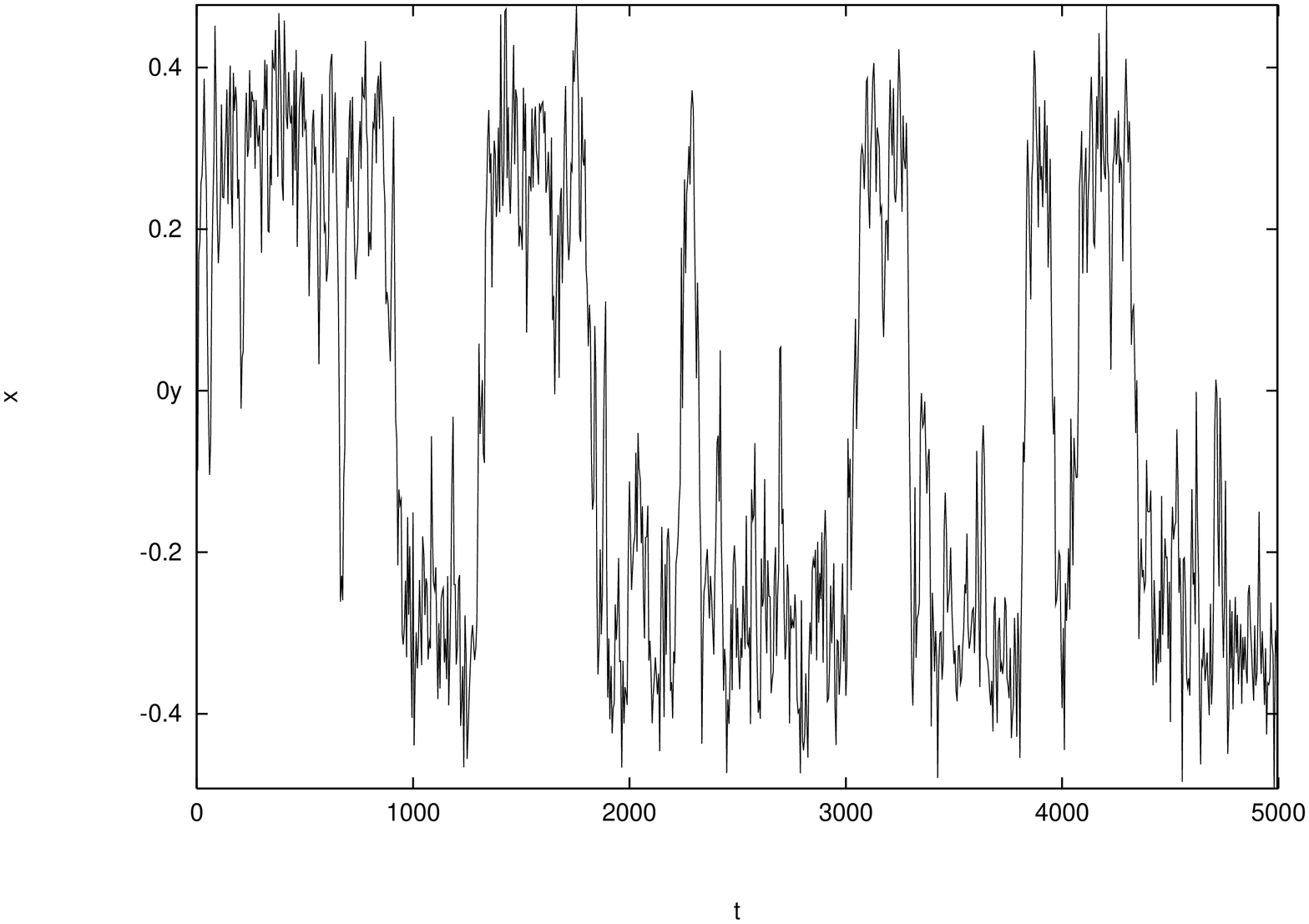}
        \vspace{0.5cm}
      \end{psfrags}
% Unterschrift
      \vspace{0.5cm}
      \caption[Calculated noisy time series]
      {{\footnotesize Variable $q$ versus time $t$. The time series is
          calculated according to the Langevin equation \\
          $\dot q(t)=0.1 q(t)-q^3(t)+0.05 \cdot F(t),$ \\
          where $F(t)$ is a Gaussian distributed fluctuation force.
          }}
      \label{fig:11}
      }
  \end{center}
\end{figure}

The first example deals with the case of one-dimensional systems,
following the Langevin equation (\ref{ex1})

\begin{equation}\label{ex1}
\dot q(t)= \epsilon q(t)-q(t)^3 +\Gamma(t)
\end{equation}

\noindent
which is valid for systems exhibiting noisy pitchfork bifurcations.
Figure \ref{fig:11} presents a part of the calculated noisy time
series.

\begin{figure}[H]
  \begin{center}
    {\begin{psfrags}
% Achsenbeschriftung
        \psfrag{D1(x)}[cc][]{{\footnotesize  Drift Coefficient
            $D^{(1)}(q)$}}
        \psfrag{x}[cc][]{{\footnotesize Variable q}}
% Diagrammbeschriftung
        \psfrag{0.1*x-x*x*x}[tr][]{{\scriptsize
            $D^{(1)}(q)=0.1q-q^3$}}
% y-Achse
        \psfrag{0}[r][]{{\scriptsize 0\,\,\,}}        
        \psfrag{0.05}[r][]{{\scriptsize 0.05}}
        \psfrag{0.1}[r][]{{\scriptsize 0.1}}
        \psfrag{-0.05}[r][]{{\scriptsize -0.05}}
        \psfrag{-0.1}[r][]{{\scriptsize -0.1}}
% x-Achse
        \psfrag{0x}[t][]{{\scriptsize 0}}        
        \psfrag{-0.4}[t][]{{\scriptsize -0.4}}
        \psfrag{-0.2}[t][]{{\scriptsize -0.2}}
        \psfrag{0.2}[t][]{{\scriptsize 0.2}}
        \psfrag{0.4}[t][]{{\scriptsize 0.4}}
% Groesse des Bildes
        \includegraphics[width=5cm,height=8.0cm,angle=270]{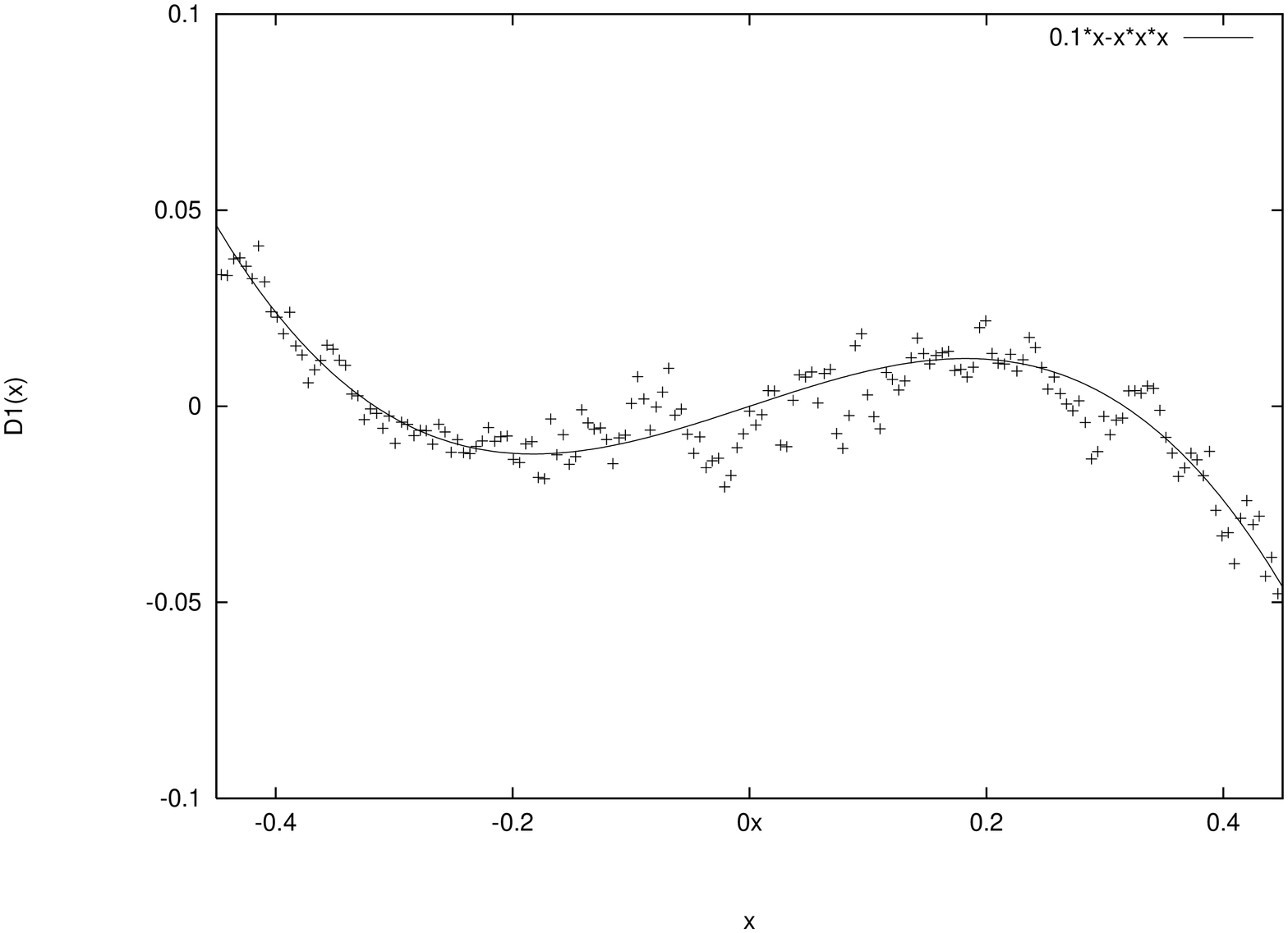}
        \vspace{0.5cm}
      \end{psfrags}
      \vspace{1cm}
% zweites Bild
      \begin{psfrags}
% Achsenbeschriftung
        \psfrag{D2(x)}[cc][]{{\footnotesize
            Diffusion Coefficient $D^{(2)}(q)$}}
        \psfrag{x}[cc][]{{\footnotesize Variable q}}
% Diagrammbeschriftung
        \psfrag{1*0.0025}[tr][]{{\scriptsize
            $D^{(2)}(q)=0.0025$}}
% y-Achse
        \psfrag{0}[r][]{{\scriptsize 0\,\,\,\,}}        
        \psfrag{0.001}[r][]{{\scriptsize  0.001}}
        \psfrag{0.002}[r][]{{\scriptsize  0.002}}
        \psfrag{0.003}[r][]{{\scriptsize  0.003}}
        \psfrag{0.004}[r][]{{\scriptsize  0.004}}
        \psfrag{0.005}[r][]{{\scriptsize  0.005}}
% x-Achse
        \psfrag{0x}[t][]{{\scriptsize 0}}
        \psfrag{-0.4}[t][]{{\scriptsize -0.4}}
        \psfrag{-0.2}[t][]{{\scriptsize -0.2}}
        \psfrag{0.2}[t][]{{\scriptsize 0.2}}
        \psfrag{0.4}[t][]{{\scriptsize 0.4}}
% Groesse des Bildes
        \includegraphics[width=5cm,height=8.0cm,angle=270,clip]%
        {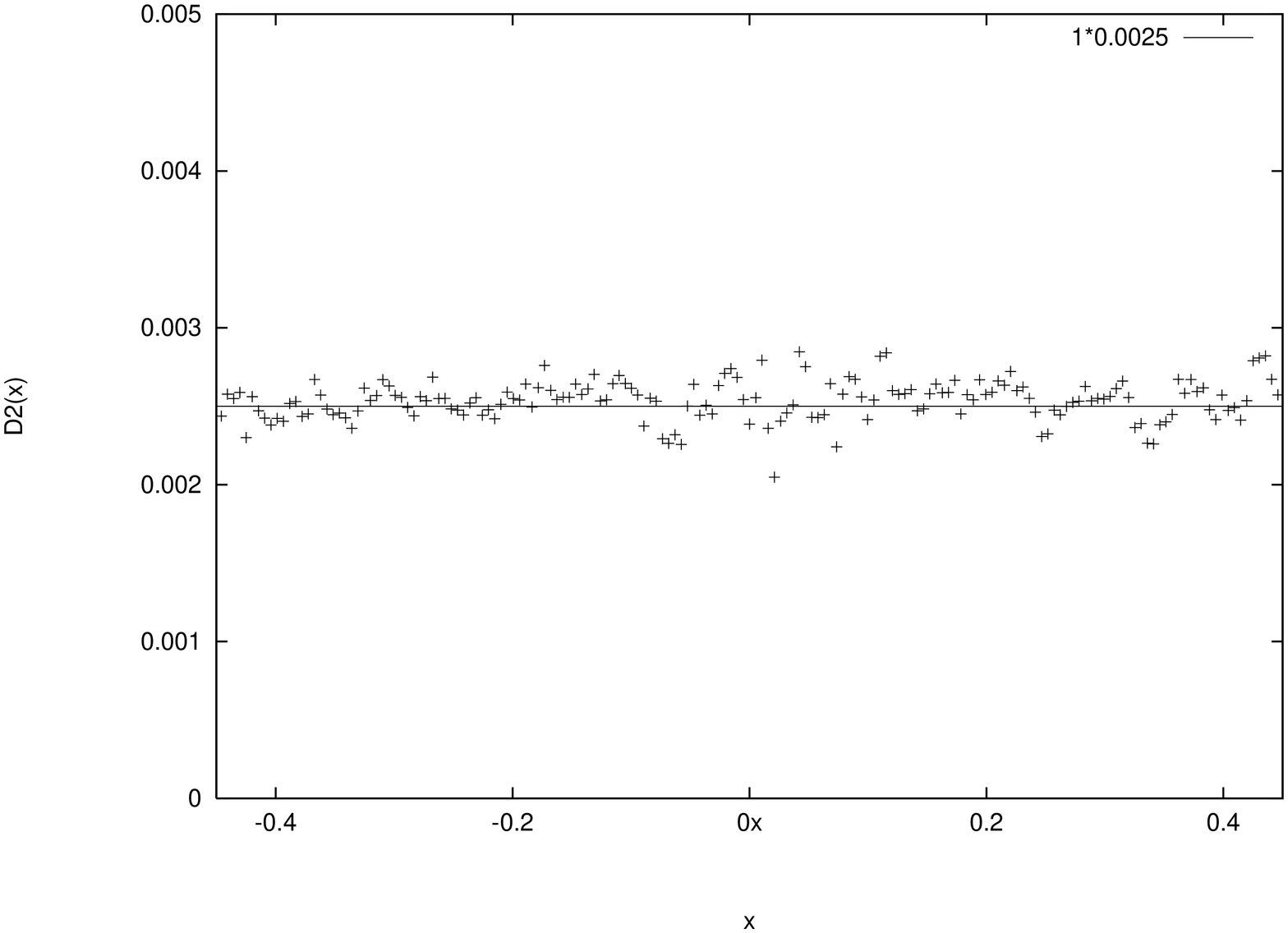}
        \vspace{0.5cm}
      \end{psfrags}
% Unterschrift
  \caption[Drift and diffusion coefficient]
  {{\footnotesize Drift and diffusion coefficient $D^{(1)}$ and
      $D^{(2)}$ versus variable $q$: According to the discussed
      algorithm the conditional probability distribution of the noisy
      time series shown in figure \ref{fig:11} has been determined and
      drift and diffusion terms have been calculated. The dots exhibit
      the numerically determined values, the lined curves show the
      theoretical functions for the coefficients according to equation
      (\ref{Drift2}) and (\ref{Diff2}). }}
\label{fig:12}
      }
  \end{center}
\end{figure}

Figure \ref{fig:12} shows drift and diffusion terms for this time
series, calculated by determining the conditional probability density
distribution for the variable q and using equations (\ref{Drift1}) and
(\ref{Diff1}).

\vspace {0.3cm}

{\centerline {\bf B. Second example}}
\vspace{0.3cm}

The second example is concerned with the case of one-dimensional
systems following the Langevin equation (\ref{ex2}).

\begin{equation}\label{ex2}
\dot \phi = \omega + \sin(\phi) + \Gamma(t).
\end{equation}

\begin{figure}[H]
  \begin{center}
    {\begin{psfrags}
% Achsenbeschriftung
        \psfrag{phi}[cc][]{{\footnotesize  Phase Difference
            $\phi (t)$}}
        \psfrag{t}[cc][]{{\footnotesize Time t}}
% y-Achse
        \psfrag{-5}[r][]{{\scriptsize -5}}
        \psfrag{0y}[r][]{{\scriptsize 0}}
        \psfrag{5}[r][]{{\scriptsize 5\,}}
        \psfrag{10}[r][]{{\scriptsize 10}}
        \psfrag{15}[r][]{{\scriptsize 15}}
        \psfrag{20}[r][]{{\scriptsize 20}}
% x-Achse
        \psfrag{0}[t][]{{\scriptsize 0}}
        \psfrag{5000}[t][]{{\scriptsize 5000}}
        \psfrag{10000}[t][]{{\scriptsize 10000}}
        \psfrag{15000}[t][]{{\scriptsize 15000}}
        \psfrag{20000}[t][]{{\scriptsize 20000}}
        \psfrag{25000}[t][]{{\scriptsize 25000}}
        \psfrag{30000}[t][]{{\scriptsize 30000}}
% Groesse des Bildes
        \includegraphics[width=5cm,height=8.0cm,angle=270]{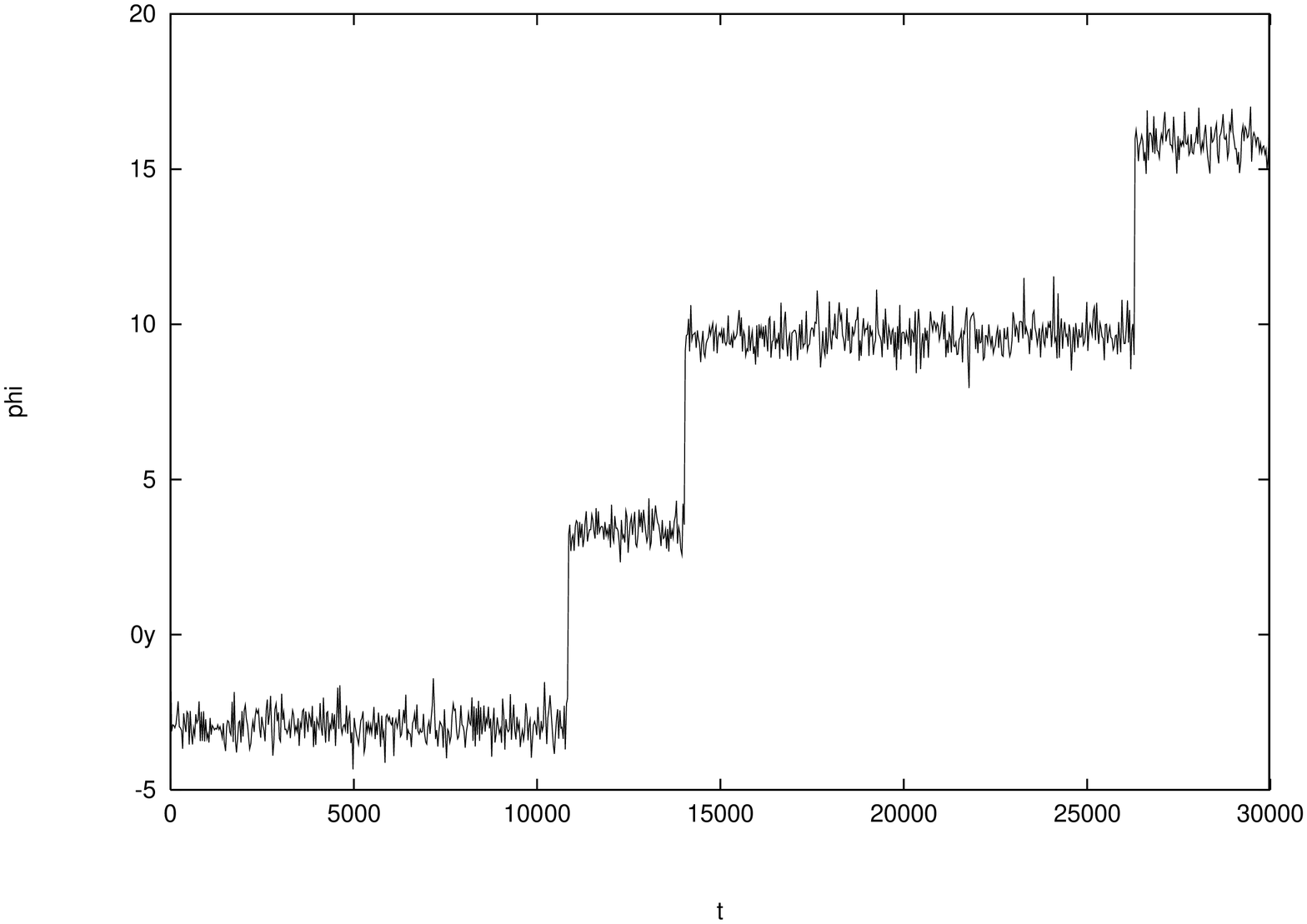}
        \vspace{0.5cm}
      \end{psfrags}
      \vspace{1cm}
% zweites Bild
      \begin{psfrags}
% Achsenbeschriftung
        \psfrag{D1(phi)}[cc][]{{\footnotesize
            Drift coefficient $D^{(1)}(\phi)$}}
        \psfrag{phi}[cc][]{{\footnotesize Phase Difference $\phi$}}
% Diagrammbeschriftung
        \psfrag{0.2+sin(x)}[tr][]{{\scriptsize
            $0.2+sin(\phi)$}}
% y-Achse
        \psfrag{0}[r][]{{\scriptsize 0}}
        \psfrag{-4}[r][]{{\scriptsize -4}}
        \psfrag{4}[r][]{{\scriptsize 4}}
        \psfrag{8}[][]{{\scriptsize 8 }}
        \psfrag{12}[][]{{\scriptsize 12\,}}
% x-Achse
        \psfrag{0x}[t][]{{\scriptsize 0}}
        \psfrag{-5}[t][]{{\scriptsize -5}}
        \psfrag{5}[t][]{{\scriptsize 5}}
        \psfrag{10}[t][]{{\scriptsize 10}}
        \psfrag{15}[t][]{{\scriptsize 15}}
% Groesse des Bildes
        \includegraphics[width=5cm,height=8.0cm,angle=270]{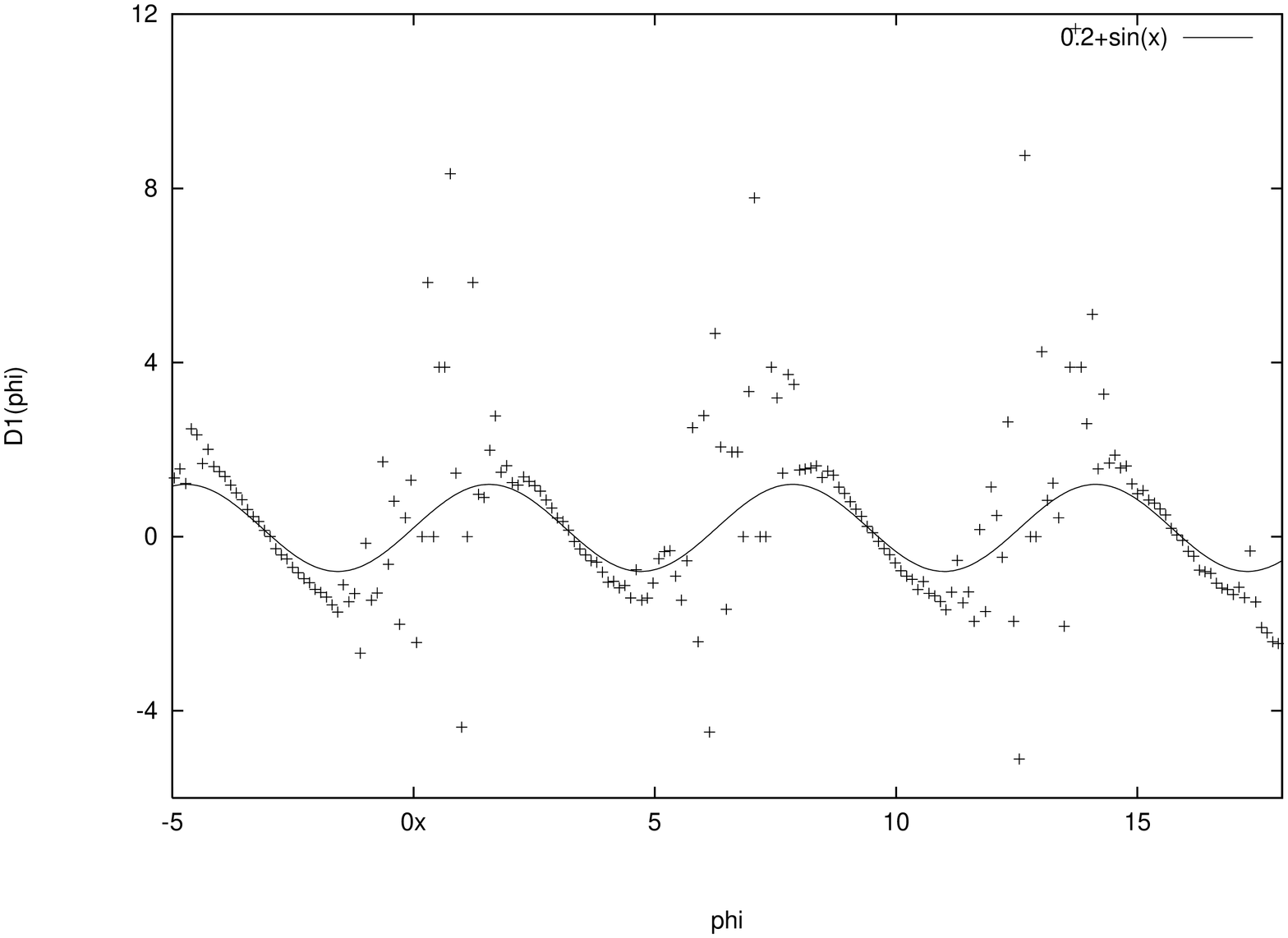}
        \vspace{0.5cm}
      \end{psfrags}
% Unterschrift
      \caption[Time series and drift coefficient]
      {{\footnotesize Phase Difference $\phi$ versus time $t$ and
          drift coefficient $D^{(1)}$ versus $\phi$: The
          noisy time series belongs to the Langevin equation\\
          $\dot\phi(t)=0.2+\sin(\phi(t))+0.6\cdot F(t)$ \\
          where F(t) is a Gaussian distributed fluctuating force. The
          corresponding drift coefficient $D^{(1)}$ has been
          calculated and is plottet (dots) together with the
          theoretical function (lined curve). }}
      \label{fig:21}
      }
  \end{center}
\end{figure}

\begin{figure}[H]
  \begin{center}
    {\begin{psfrags}
% Achsenbeschriftung
        \psfrag{phi}[cc][]{{\footnotesize  Phase Difference
            $\phi (t)$}}
        \psfrag{t}[cc][]{{\footnotesize Time t}}
% y-Achse
        \psfrag{0y}[][]{{\scriptsize 0}}
        \psfrag{-2}[r][]{{\scriptsize -2}}
        \psfrag{2}[r][]{{\scriptsize 2}}
        \psfrag{4}[r][]{{\scriptsize 4}}
        \psfrag{6}[r][]{{\scriptsize 6}}
        \psfrag{8}[r][]{{\scriptsize 8}}
        \psfrag{10}[][]{{\scriptsize 10\,\,}}
        \psfrag{12}[][]{{\scriptsize 12}}
% x-Achse
        \psfrag{0}[t][]{{\scriptsize 0}}
        \psfrag{20}[t][]{{\scriptsize 20}}
        \psfrag{40}[t][]{{\scriptsize 40}}
        \psfrag{60}[t][]{{\scriptsize 60}}
        \psfrag{80}[t][]{{\scriptsize 80}}
        \psfrag{100}[t][]{{\scriptsize 100}}
% Groesse des Bildes
        \includegraphics[width=5cm,height=8.0cm,angle=270]{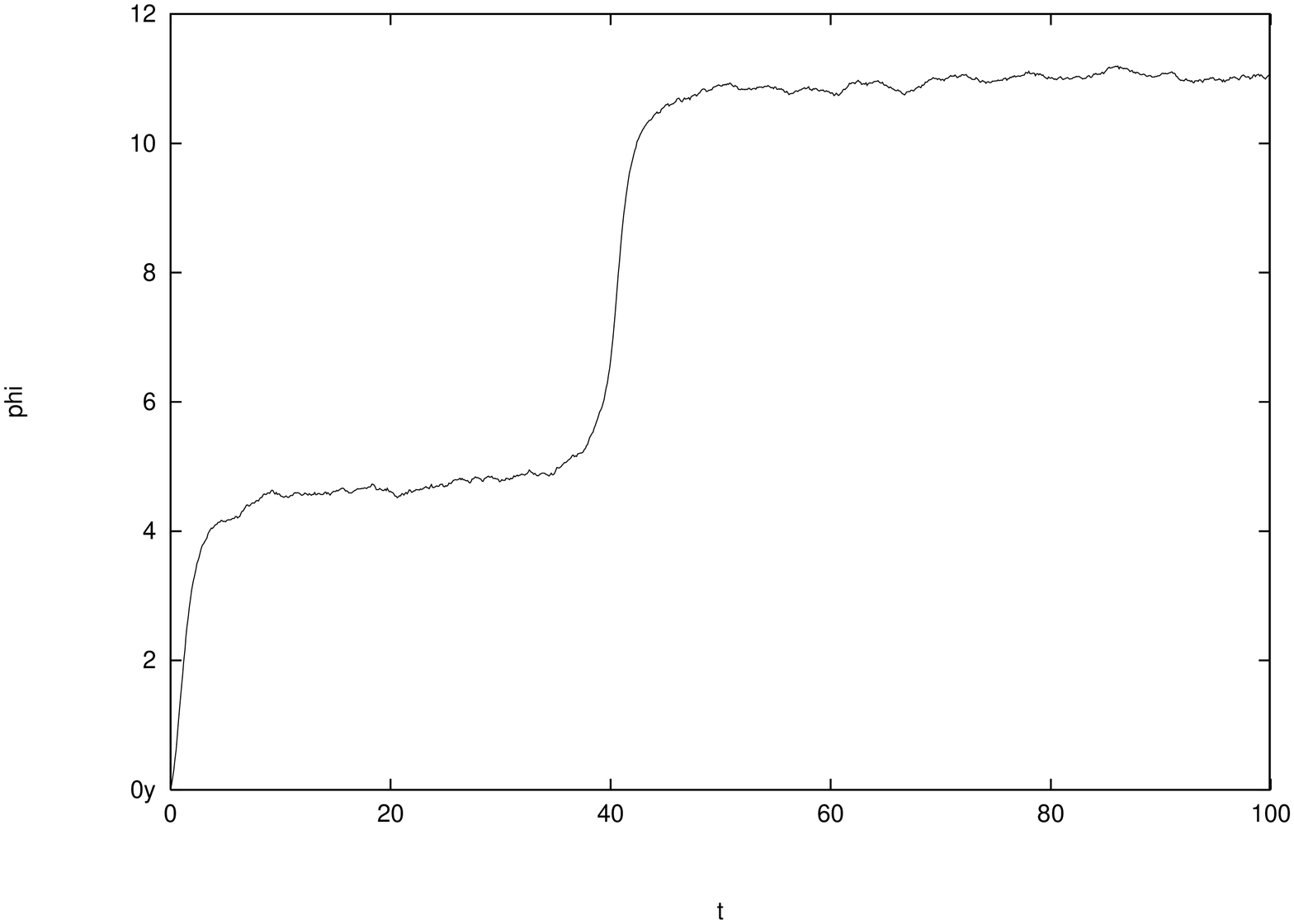}
        \vspace{0.5cm}
      \end{psfrags}
      \vspace{1cm}
% zweites Bild
      \begin{psfrags}
% Achsenbeschriftung
        \psfrag{D1(phi)}[cc][]{{\footnotesize
            Drift coefficient $D^{(1)}(\phi)$}}
        \psfrag{phi}[cc][]{{\footnotesize Phase Difference $\phi$}}
% Diagrammbeschriftung
        \psfrag{1.0+sin(x)}[tr][]{{\scriptsize
            $1.0+sin(\phi)$}}
% y-Achse
        \psfrag{0y}[][]{{\scriptsize 0\,\,\,}}
        \psfrag{-4}[r][]{{\scriptsize -4}}
        \psfrag{-2y}[r][]{{\scriptsize -2}}
        \psfrag{2y}[r][]{{\scriptsize 2}}
        \psfrag{4y}[r][]{{\scriptsize 4}}
% x-Achse
        \psfrag{-2}[t][]{{\scriptsize -2}}
        \psfrag{0}[t][]{{\scriptsize 0}}
        \psfrag{2}[t][]{{\scriptsize 2}}
        \psfrag{4}[t][]{{\scriptsize 4}}
        \psfrag{6}[t][]{{\scriptsize 6}}
        \psfrag{8}[t][]{{\scriptsize 8}}
        \psfrag{10}[t][]{{\scriptsize 10}}
        \psfrag{12}[t][]{{\scriptsize 12}}
% Groesse des Bildes
        \includegraphics[width=5cm,height=8.0cm,angle=270]{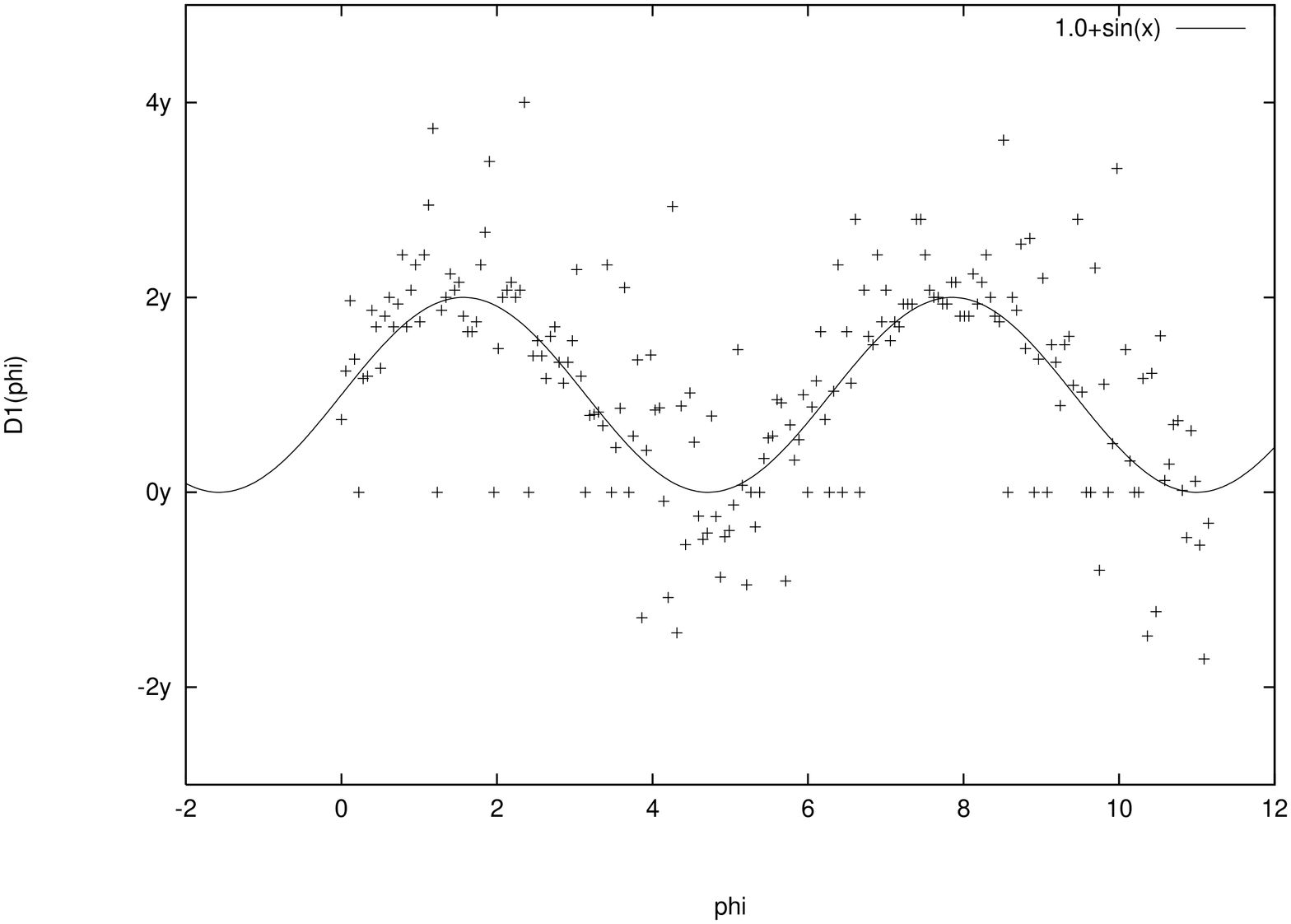}
        \vspace{0.5cm}
      \end{psfrags}
% Unterschrift
      \caption[Time series and drift coefficient]
      {{\footnotesize Phase Difference $\phi$ versus time $t$ and
          drift coefficient $D^{(1)}$ versus $\phi$: The
          noisy time series belongs to the Langevin equation\\
          $\dot\phi(t)=1.0+\sin(\phi(t))+0.05\cdot F(t)$ \\
          where F(t) is a Gaussian distributed fluctuating force. In
          the second part of the figure the numerically determined
          corresponding drift coefficient $D^{(1)}$ is plottet
          together with the theoretical function (lined curve).  }}
      \label{fig:31}
      }
  \end{center}
\end{figure}

This type of equation describes the dynamics of a phase difference
$\phi=\phi_1-\phi_2$, where $\phi_1,\, \phi_2$ are the phases of two
coupled nonlinear oscillators. For two different sets of parameters
the noisy time series and calculated drift coefficients are shown in
fig. \ref{fig:21} and \ref{fig:31}. Because of the singularities of
the time series there is only a small probability for the
corresponding variable values. These rapid changes of the phase are
responsible for the strong noise of the drift coefficient.

Since the Langevin equation (\ref{ex2}) is valid for a wide class of
biological systems applications of the algorithm for many different
fields are expected.

\vspace {0.3cm}

{\centerline {\bf C. Third example}}
\vspace{0.3cm}

After these two examples of one-dimensional systems the case of two
variables will be discussed. In the following examples the presented
method will be applied on vectors with two components. The third
example is based on the differential equation system of a Hopf
bifurcation:

\begin{eqnarray}
  \label{Hopf1}
  \frac{d}{dt}q_1 & = &\epsilon q_1
  - \gamma q_2+(q_1^2+q_2^2)(\mu q_1-\omega q_2) \\
  \label{Hopf2}
  \frac{d}{dt}q_2& = &\gamma q_1
  - \epsilon q_2+(q_1^2+q_2^2)(\omega q_1+\mu q_2)
\end{eqnarray}

\begin{figure}[H]
  \begin{center}
    {\begin{psfrags}
% Achsenbeschriftung
        \psfrag{xachse}[cc][]{{\footnotesize Variable $q_1$}}
        \psfrag{yachse}[cc][]{{\footnotesize Variable $q_2$}}
% x-Achse
        \psfrag{-0.15x}[t][]{{\scriptsize -0.15}}
        \psfrag{-0.1x}[t][]{{\scriptsize -0.1}}
        \psfrag{-0.05x}[t][]{{\scriptsize -0.05}}
        \psfrag{0x}[t][]{{\scriptsize 0}}
        \psfrag{0.05x}[t][]{{\scriptsize 0.05}}
        \psfrag{0.1x}[t][]{{\scriptsize 0.1}}
        \psfrag{0.15x}[t][]{{\scriptsize 0.15}}
% y-Achse
        \psfrag{-0.15}[r][]{{\scriptsize -0.15}}
        \psfrag{-0.1}[r][]{{\scriptsize -0.1}}
        \psfrag{-0.05}[r][]{{\scriptsize -0.05}}
        \psfrag{0}[r][]{{\scriptsize 0\,\,}}
        \psfrag{0.05}[r][]{{\scriptsize 0.05}}
        \psfrag{0.1}[r][]{{\scriptsize 0.1}}
        \psfrag{0.15}[r][]{{\scriptsize 0.15}}
% Groesse des Bildes
        \includegraphics[width=5.8cm,height=6cm,angle=270]{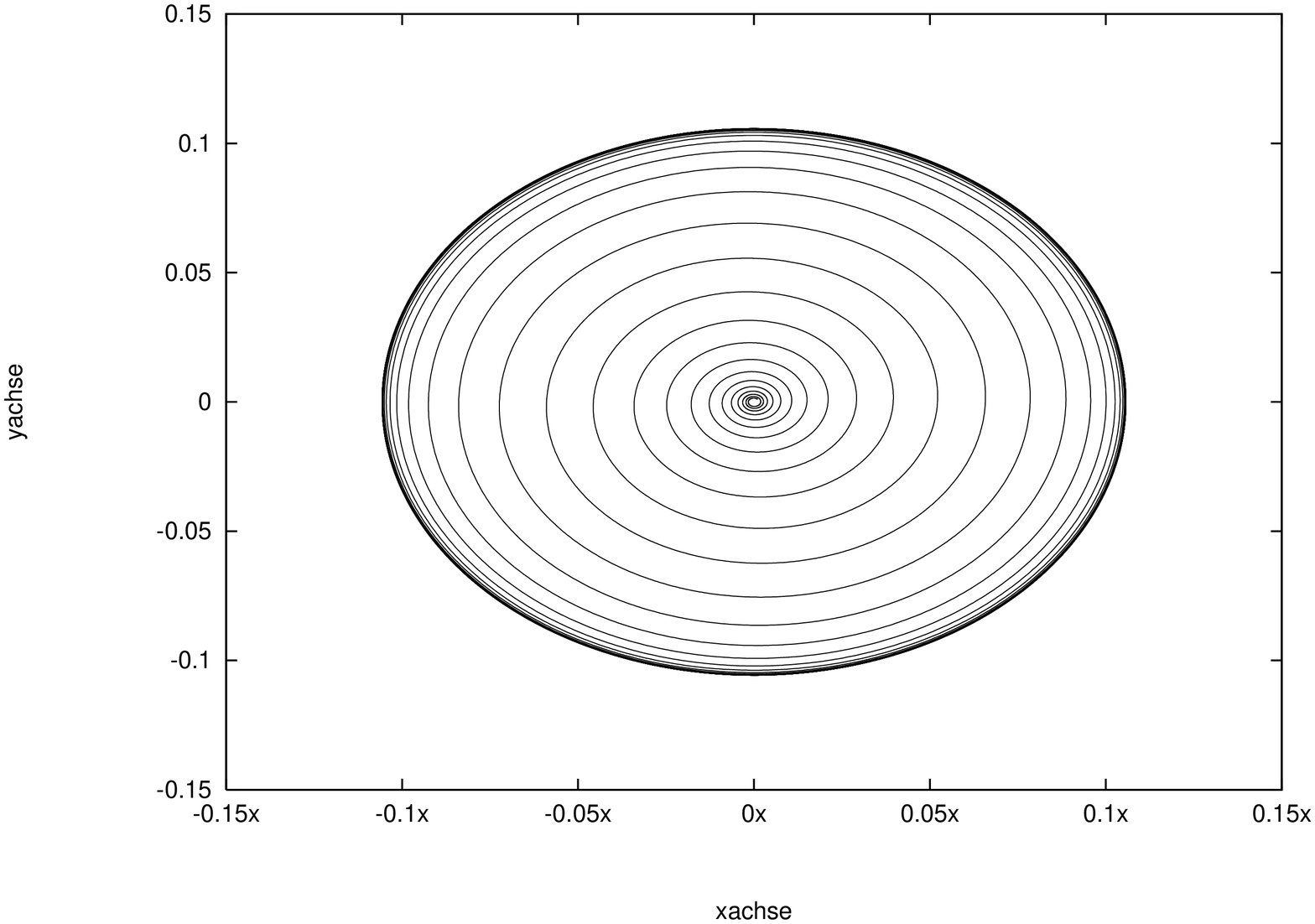}
        \vspace{0.5cm}
      \end{psfrags}
% Unterschrift
      \caption[Phase diagram]{
        {\footnotesize Variable $q_2$ versus variable $q_1$: The phase
          diagram shows the dynamics of a Hopf
          bifurcation:\\
          $\dot q_1=0.05q_1-q_2+(q_1^2+q_2^2)(-5q_1-7.5q_2)$ \\
          $\dot q_2=q_1-0.05q_2+(q_1^2+q_2^2)(7.5q_1-5q_2)$ }}
      \label{fig:51}
      }
  \end{center}
\end{figure}

Fig. \ref{fig:51} shows the phase diagram of the $q_1$ and $q_2$. The
parameters

\begin{equation}
  \epsilon =0.05, \qquad \gamma =1, \qquad \mu =-5, \qquad \omega
  =7.5
\end{equation}

are chosen in a way that the behaviour of the variables becomes
supercritical, i.e. the focus (0,0) is instable and the trajectory
moves towards a stable limit cycle \cite{Uhl}. This system of
differential equations is very important for describing time-spatial
signals whose dynamics are determined by two order parameters.

\begin{figure}[H]
  \begin{center}
    {\begin{psfrags}
% Achsenbeschriftung
        \psfrag{xachse}[cc][]{{\footnotesize Variable $q_1$}}
        \psfrag{yachse}[tl][]{{\footnotesize Variable $q_2$}}
        \psfrag{zachse}[cc][]{{\footnotesize
            Drift Coeff. $D^{(1)}_{q_1}({\bf q})$}}
% z-Achse
        \psfrag{-4z}[r][]{{\scriptsize -4}}
        \psfrag{0z}[r][]{{\scriptsize 0}}
        \psfrag{4z}[r][]{{\scriptsize 4}}
        \psfrag{8z}[r][]{{\scriptsize 8}}
% x-Achse
        \psfrag{-0.8x}[t][]{{\scriptsize -0.8 }}
        \psfrag{0x}[t][]{{\scriptsize 0}}
        \psfrag{-0.4x}[t][]{{\scriptsize -0.4}}
        \psfrag{0.4x}[t][]{{\scriptsize 0.4}}
% y-Achse
        \psfrag{-0.8}[t][]{{\scriptsize -0.8 }}
        \psfrag{-0.4}[t][]{{\scriptsize -0.4}}
        \psfrag{0.4}[t][]{{\scriptsize 0.4}}
        \psfrag{0}[t][]{{\scriptsize 0}}
% Groesse des Bildes
        \includegraphics[width=5cm,height=8.0cm,angle=270]{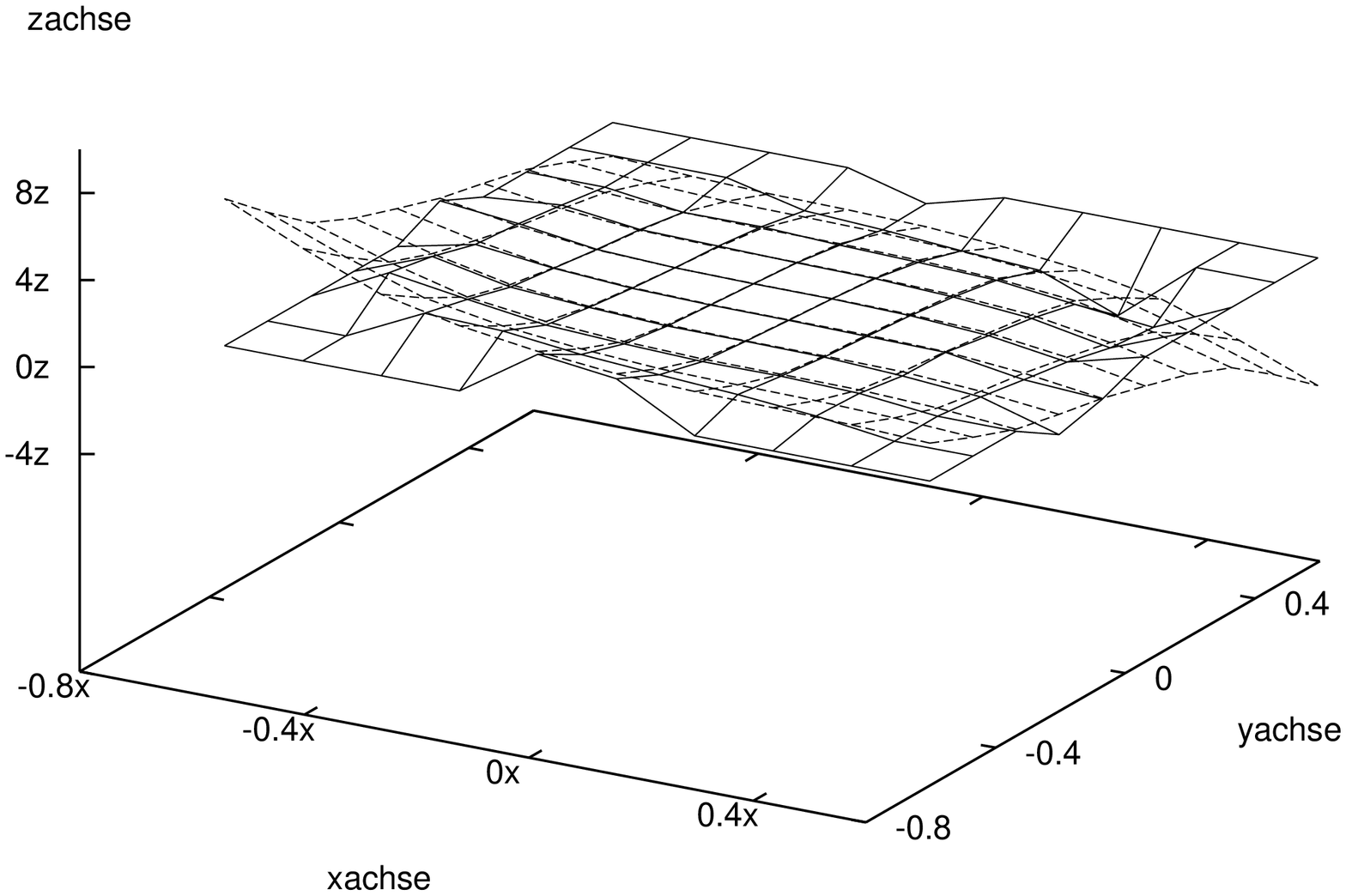}
        \vspace{0.5cm}
      \end{psfrags}
      \vspace{1cm}
% zweites Bild
      \begin{psfrags}
% Achsenbeschriftung
        \psfrag{xachse}[cc][]{{\footnotesize Variable $q_1$}}
        \psfrag{yachse}[tl][]{{\footnotesize Variable $q_2$}}
        \psfrag{zachse}[cc][]{{\footnotesize
            Drift Coeff. $D^{(1)}_{q_2}({\bf q})$}}
% Nullpunkt
        \psfrag{0}[][]{{\scriptsize 0}}        
% z-Achse
        \psfrag{200z}[r][]{{\scriptsize 200}}
        \psfrag{100z}[r][]{{\scriptsize 100}}
        \psfrag{0z}[r][]{{\scriptsize 0\,\,\,}}
        \psfrag{-100z}[r][]{{\scriptsize -100}}
        \psfrag{-200z}[r][]{{\scriptsize -200}}
% x-Achse
        \psfrag{-0.8x}[t][]{{\scriptsize -0.8 }}
        \psfrag{0x}[t][]{{\scriptsize 0}}
        \psfrag{-0.4x}[t][]{{\scriptsize -0.4}}
        \psfrag{0.4x}[t][]{{\scriptsize 0.4}}
% y-Achse
        \psfrag{-0.8}[t][]{{\scriptsize -0.8 }}
        \psfrag{-0.4}[t][]{{\scriptsize -0.4}}
        \psfrag{0.4}[t][]{{\scriptsize 0.4}}
        \psfrag{0}[t][]{{\scriptsize 0}}
% Groesse des Bildes
        \includegraphics[width=5cm,height=8.0cm,angle=270]{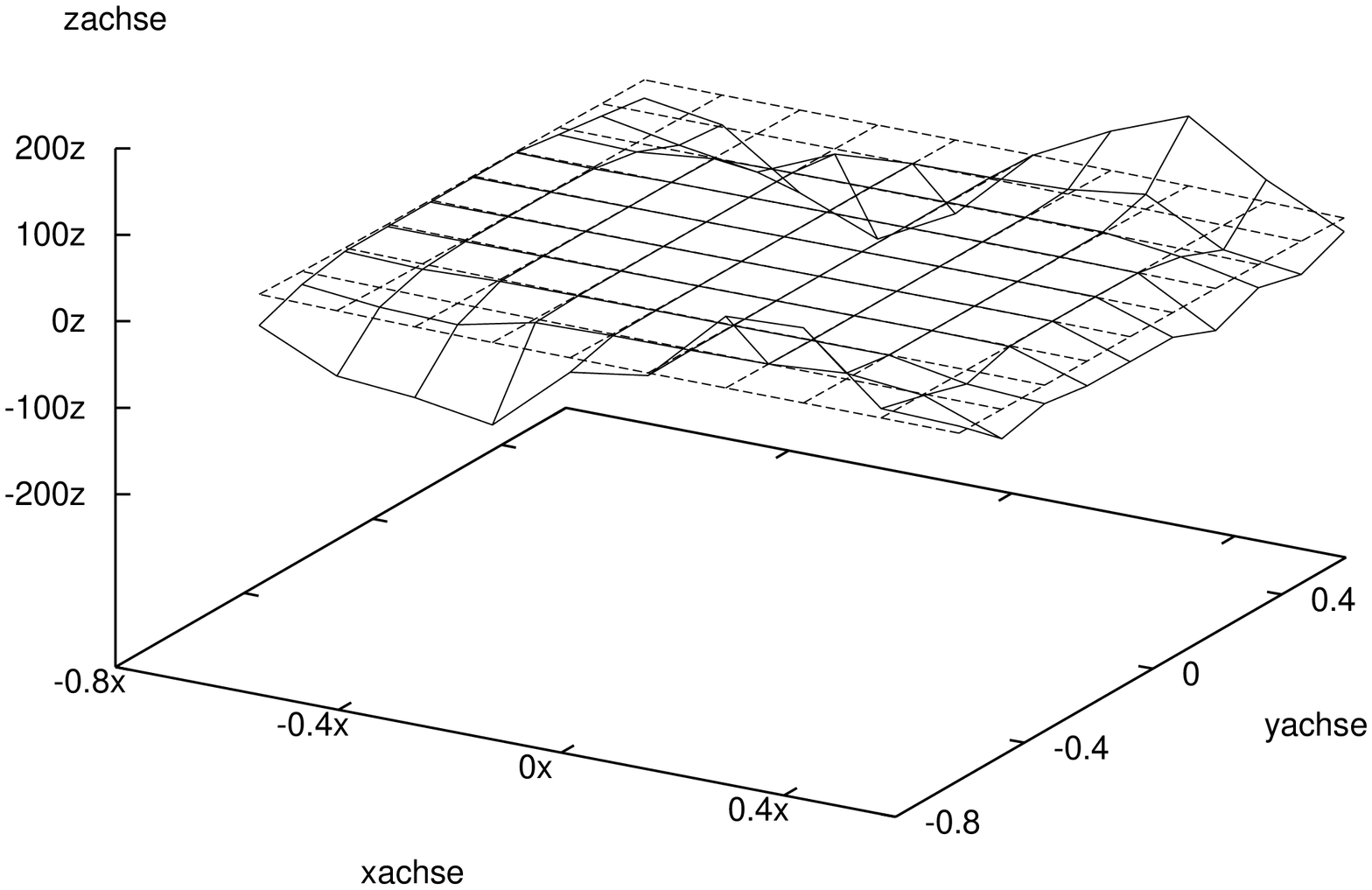}
        \vspace{0.5cm}
      \end{psfrags}
% Unterschrift
      \caption[Drift coefficients]{
      {\footnotesize Drift coefficients $D_{q_1}^{(1)}$ and
          $D_{q_2}^{(1)}$ versus variables $q_1$ and $q_2$:
          The lined surfaces belong to the calculated data, the dashed
          surfaces are plottet according to the expected surfaces:\\
          $D_{q_1}{(1)}({\bf q})=0.05 q_1-q_2+(q_1^2+q_2^2)
          (-5q_1-7.5q_2)$ \\
          $D_{q_2}{(1)}({\bf q})=q_1-0.05 q_2+(q_1^2+q_2^2)
          (7.5q_1-5q_2)$
          }}
      \label{fig:41}
      }
  \end{center}
\end{figure}

A Gaussian distributed white noise weighted by a factor 0.2 has been
added to the system described by equations (\ref{Hopf1}) and
(\ref{Hopf2}). Afterwards the presented method has been applied.  Fig.
\ref{fig:41} shows the two numerically determined drift coefficients
$D_{q_1}^{(1)}({\bf q})$ and $D_{q_2}^{(1)}({\bf q})$.  The lined
surface belongs to the calculated terms, the dashed one to the
expected surface. The differences at the borders are caused by too few
variable values in these regions. These values are taken only from a
few stochastic realizations, so that the algorithm cannot work with
great accuracy.  In the main field a good conformance between
calculated and expected surfaces can be seen.

\vspace {0.3cm}

{\centerline {\bf D. Fourth example}}
\vspace{0.3cm}

The fourth example deals with the case of a co-dimen\-sion~II system.
The dynamics of such a system is determined by the differential
equations (\ref{II1}) and (\ref{II2})

\begin{eqnarray}
  \label{II1}
  \frac{d}{dt}q_1&=&q_2 + F_1(t) \\
  \label{II2}
  \frac{d}{dt}q_2&=&\epsilon q_1+ \gamma q_2 + \mu q_1^3 + \omega
  q_1^2q_2 + F_2(t)
\end{eqnarray}

The parameters have been chosen as

\begin{equation}
  \label{II3}
  \epsilon = 0.01, \qquad \gamma = 0.03, \qquad \mu = -1,
  \qquad \omega = -1. 
\end{equation}

The fluctuation force $F(t)$ has been taken as a Gaussian distributed
noise weigh\-ted by a factor 0.05.

\begin{figure}[H]
  \begin{center}
    {\begin{psfrags}
% Achsenbeschriftung
        \psfrag{xachse}[cc][]{{\footnotesize Variable $q_1$}}
        \psfrag{yachse}[cc][]{{\footnotesize Variable $q_2$}}
% y-Achse
        \psfrag{-0.04}[r][]{{\scriptsize -0.04}}
        \psfrag{-0.08}[r][]{{\scriptsize -0.08}}
        \psfrag{0}[r][]{{\scriptsize 0\,\,\,}}
        \psfrag{0.04}[r][]{{\scriptsize 0.04}}
        \psfrag{0.08}[r][]{{\scriptsize 0.08}}
% x-Achse
        \psfrag{0x}[t][]{{\scriptsize 0}}
        \psfrag{-0.2x}[t][]{{\scriptsize -0.2}}
        \psfrag{-0.4x}[t][]{{\scriptsize 0.4}}
        \psfrag{0.2x}[t][]{{\scriptsize  0.2}}
        \psfrag{0.4x}[t][]{{\scriptsize 0.4}}
% Groesse des Bildes
        \includegraphics[width=5cm,height=8.0cm,angle=270]{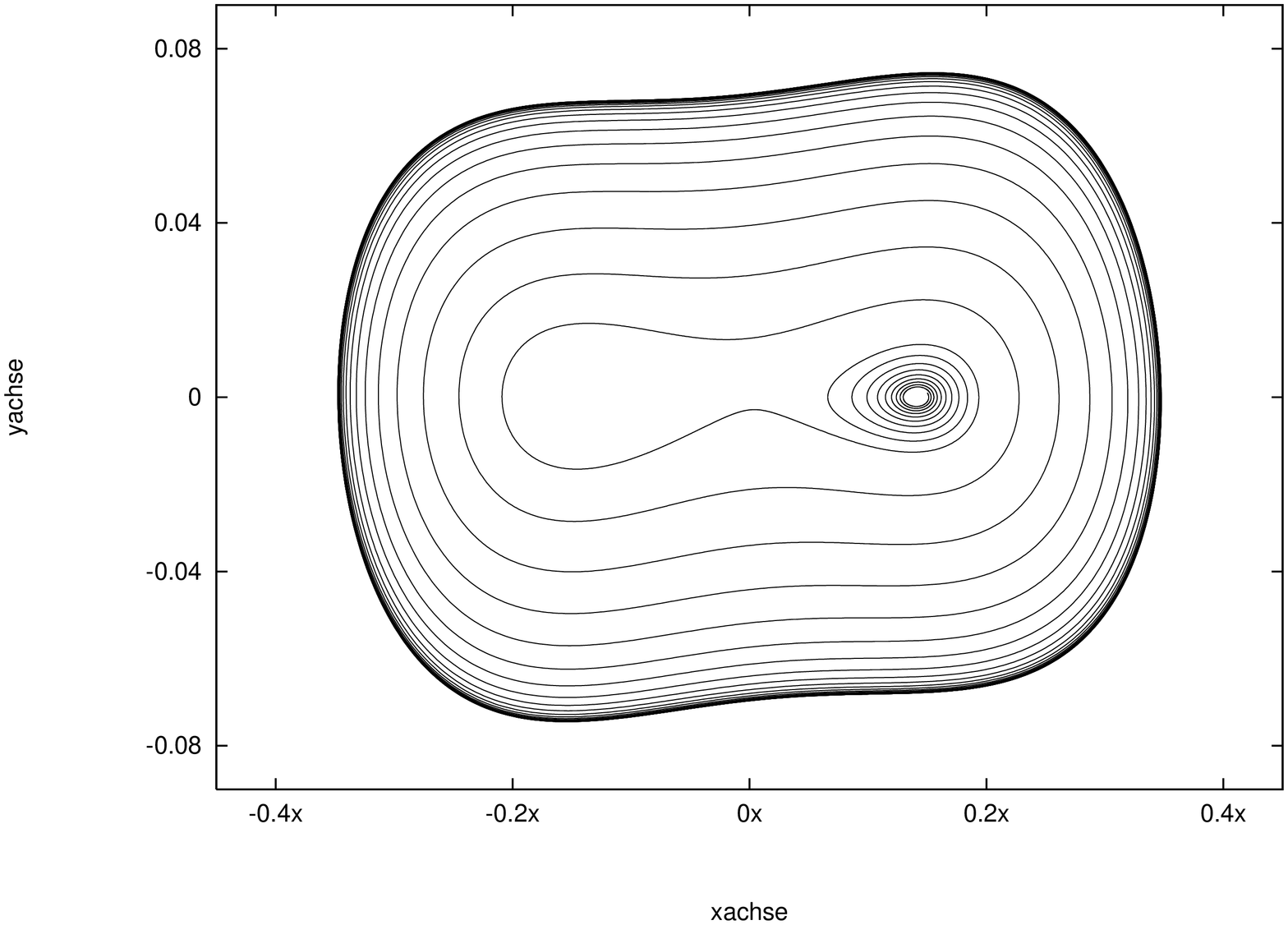}
        \vspace{0.5cm}
      \end{psfrags}
% Unterschrift
      \caption[Phase diagram]{
        {\footnotesize Variable $q_2$ versus variable $q_1$: The phase
          diagram shows the dynamics of a co-di\-men\-sion~II
          instability:\\
          $\dot q_1=q_2$  \\
          $\dot q_2=0.02 q_1+ 0.03 q_2 - q_1^3 - q_1^2q_2$ }}
      \label{fig:71}
      }
  \end{center}
\end{figure}

Fig. \ref{fig:71} shows the phase diagram for this system without a
fluctuating force $F(t)$. The trajectory is repelled from the instable
focus and moves towards a limit cycle.

\begin{figure}[H]
  \begin{center}
    {\begin{psfrags}
% Achsenbeschriftung
        \psfrag{xachse}[cc][]{{\footnotesize $q_1$}}
        \psfrag{yachse}[tl][]{{\footnotesize $q_2$}}
        \psfrag{zachse}[cc][]{{\footnotesize
            Drift coeff. $D^{(1)}_{q_1}({\bf q})$}}
% x-Achse
        \psfrag{0.4x}[t][]{{\scriptsize 0.4}}
        \psfrag{0.8x}[t][]{{\scriptsize 0.8}}
        \psfrag{-0.4x}[t][]{{\scriptsize -0.4}}
        \psfrag{-0.8x}[t][]{{\scriptsize -0.8}}
        \psfrag{0x}[t][]{{\scriptsize 0}}
% y-Achse
        \psfrag{0.2}[t][]{{\scriptsize 0.2}}
        \psfrag{0.4}[t][]{{\scriptsize 0.4}}
        \psfrag{-0.6}[t][]{{\scriptsize -0.6}}
        \psfrag{-0.4}[t][]{{\scriptsize -0.4}}
        \psfrag{-0.2}[t][]{{\scriptsize -0.2}}
        \psfrag{0}[t][]{{\scriptsize 0}}
% z-Achse
        \psfrag{0.4z}[r][]{{\scriptsize 0.4}}
        \psfrag{0.8z}[r][]{{\scriptsize 0.8}}
        \psfrag{1.2z}[r][]{{\scriptsize 1.2}}
        \psfrag{1.6z}[r][]{{\scriptsize 1.6}}
        \psfrag{-0.8z}[r][]{{\scriptsize -0.8}}
        \psfrag{-0.4z}[r][]{{\scriptsize -0.4}}
        \psfrag{0z}[r][]{{\scriptsize 0\,\,}}
% Groesse des Bildes
        \includegraphics[width=5cm,height=8.0cm,angle=270]{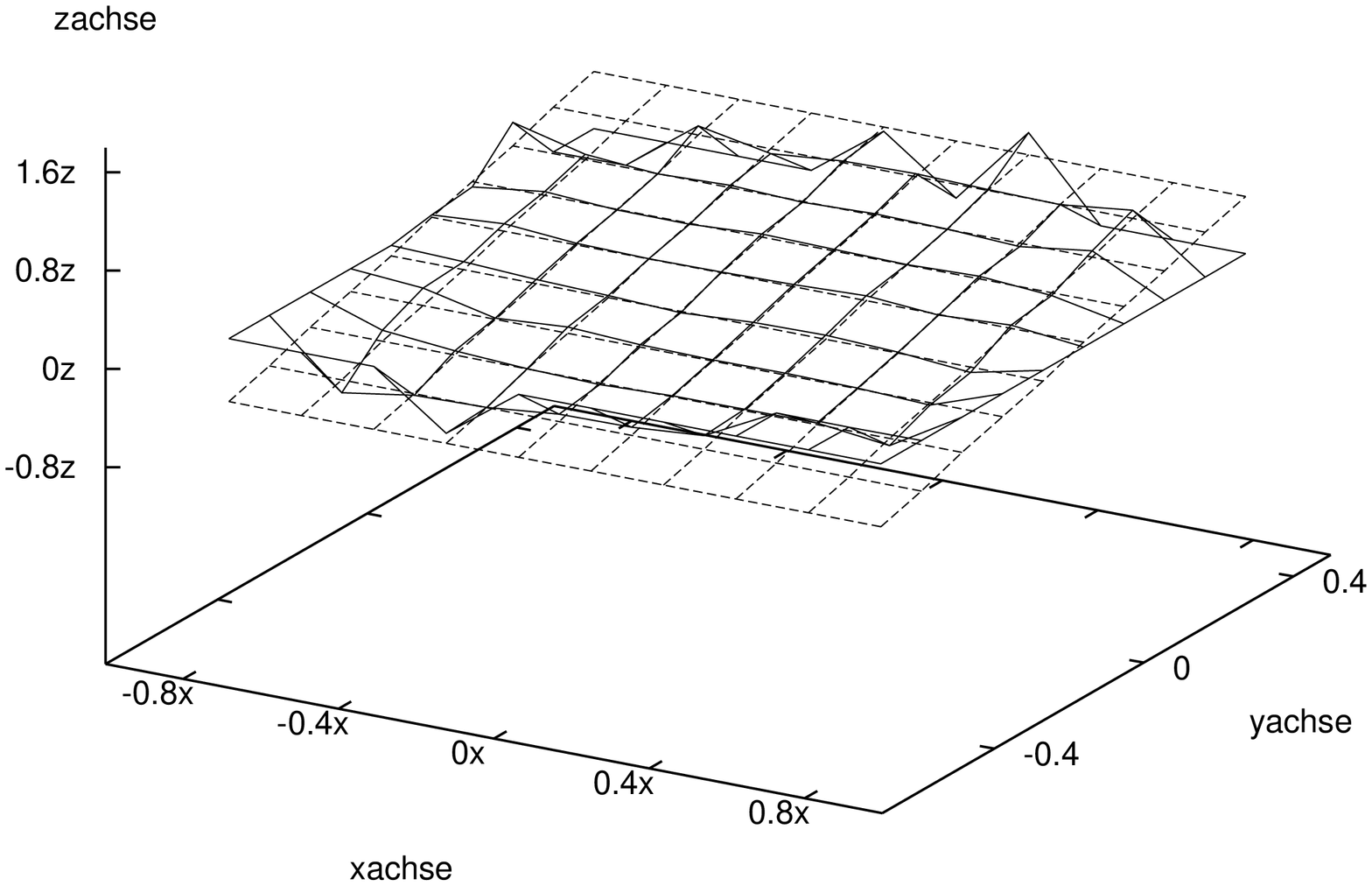}
        \vspace{0.5cm}
      \end{psfrags}
      \vspace{1cm}
% zweites Bild
      \begin{psfrags}
% Achsenbeschriftung
        \psfrag{xachse}[cc][]{{\footnotesize $q_1$}}
        \psfrag{yachse}[tl][]{{\footnotesize $q_2$}}
        \psfrag{zachse}[cc][]{{\footnotesize
            Drift coeff. $D^{(1)}_{q_2}({\bf q})$}}
% x-Achse
        \psfrag{0.4x}[t][]{{\scriptsize 0.4}}
        \psfrag{0.8x}[t][]{{\scriptsize 0.8}}
        \psfrag{-0.4x}[t][]{{\scriptsize -0.4}}
        \psfrag{-0.8x}[t][]{{\scriptsize -0.8}}
        \psfrag{0x}[t][]{{\scriptsize 0}}
% y-Achse
        \psfrag{0.2}[t][]{{\scriptsize 0.2}}
        \psfrag{0.4}[t][]{{\scriptsize 0.4}}
        \psfrag{-0.6}[t][]{{\scriptsize -0.6}}
        \psfrag{-0.4}[t][]{{\scriptsize -0.4}}
        \psfrag{-0.2}[t][]{{\scriptsize -0.2}}
        \psfrag{0}[t][]{{\scriptsize 0}}
% z-Achse
        \psfrag{0.4z}[r][]{{\scriptsize 0.4}}
        \psfrag{0.8z}[r][]{{\scriptsize 0.8}}
        \psfrag{1.2z}[r][]{{\scriptsize 1.2}}
        \psfrag{1.6z}[r][]{{\scriptsize 1.6}}
        \psfrag{-0.8z}[r][]{{\scriptsize -0.8}}
        \psfrag{-0.4z}[r][]{{\scriptsize -0.4}}
        \psfrag{0z}[r][]{{\scriptsize 0\,\,}}
% Groesse des Bildes
        \includegraphics[width=5cm,height=8.0cm,angle=270]{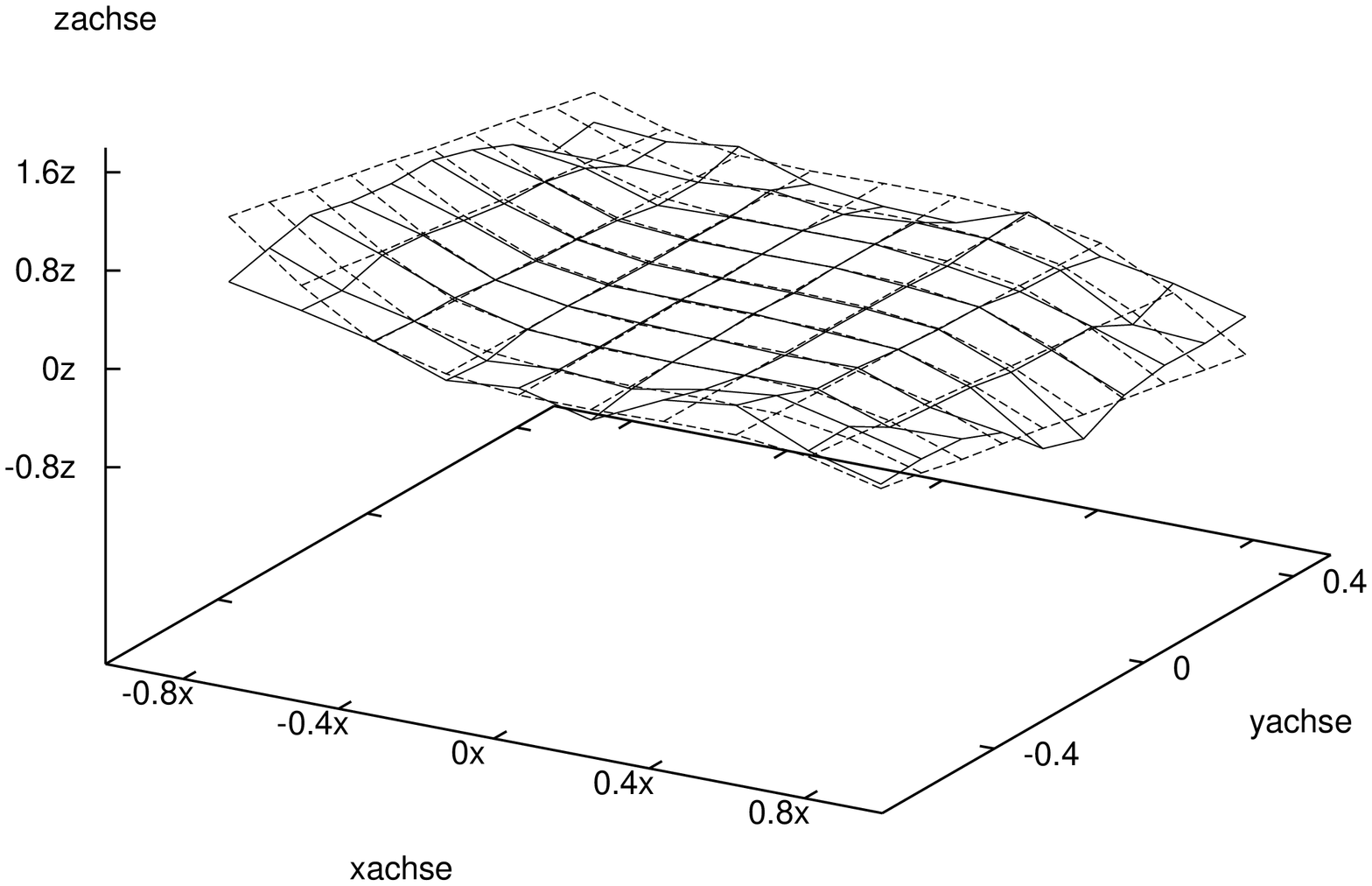}
        \vspace{0.5cm}
      \end{psfrags}
% Unterschrift
      \caption[Drift coefficients]{
      {\footnotesize Drift coefficients $D_{q_1}^{(1)}$ and
          $D_{q_2}^{(1)}$ versus variables $q_1$ and $q_2$:
          The lined surfaces belong to the calculated data, the dashed
          surfaces are drawn according to the expected surfaces:\\
            $D_{q_1}^{(1)}({\bf q})=q_2$ \\
            $D_{q_2}^{(1)}({\bf q})=0.02q_1+0.03q_2-q_1^3-q_1^2q_2$
          }}
      \label{fig:61}
      }
  \end{center}
\end{figure}

\begin{figure}[H]
  \begin{center}
    {\begin{psfrags}
% Achsenbeschriftung
        \psfrag{xachse}[cc][]{{\footnotesize $q_1$}}
        \psfrag{yachse}[tl][]{{\footnotesize $q_2$}}
        \psfrag{zachse}[cc][]{{\footnotesize Diff. coeff.
            $D^{(2)}_{q_1q_2}({\bf q})$}}
% x-Achse
        \psfrag{0.4x}[t][]{{\scriptsize 0.4}}
        \psfrag{0.8x}[t][]{{\scriptsize 0.8}}
        \psfrag{-0.4x}[t][]{{\scriptsize -0.4}}
        \psfrag{-0.8x}[t][]{{\scriptsize -0.8}}
        \psfrag{0x}[t][]{{\scriptsize 0}}
% y-Achse
        \psfrag{0.2}[t][]{{\scriptsize 0.2}}
        \psfrag{0.4}[t][]{{\scriptsize 0.4}}
        \psfrag{-0.6}[t][]{{\scriptsize -0.6}}
        \psfrag{-0.4}[t][]{{\scriptsize -0.4}}
        \psfrag{-0.2}[t][]{{\scriptsize -0.2}}
        \psfrag{0}[t][]{{\scriptsize 0}}
% z-Achse
        \psfrag{20z}[r][]{{\scriptsize 20}}
        \psfrag{10z}[r][]{{\scriptsize 10}}
        \psfrag{-10z}[r][]{{\scriptsize -10}}
        \psfrag{-20z}[r][]{{\scriptsize -20}}
        \psfrag{0z}[r][]{{\scriptsize 0}}
% Groesse des Bildes
        \includegraphics[width=5cm,height=8.0cm,angle=270]{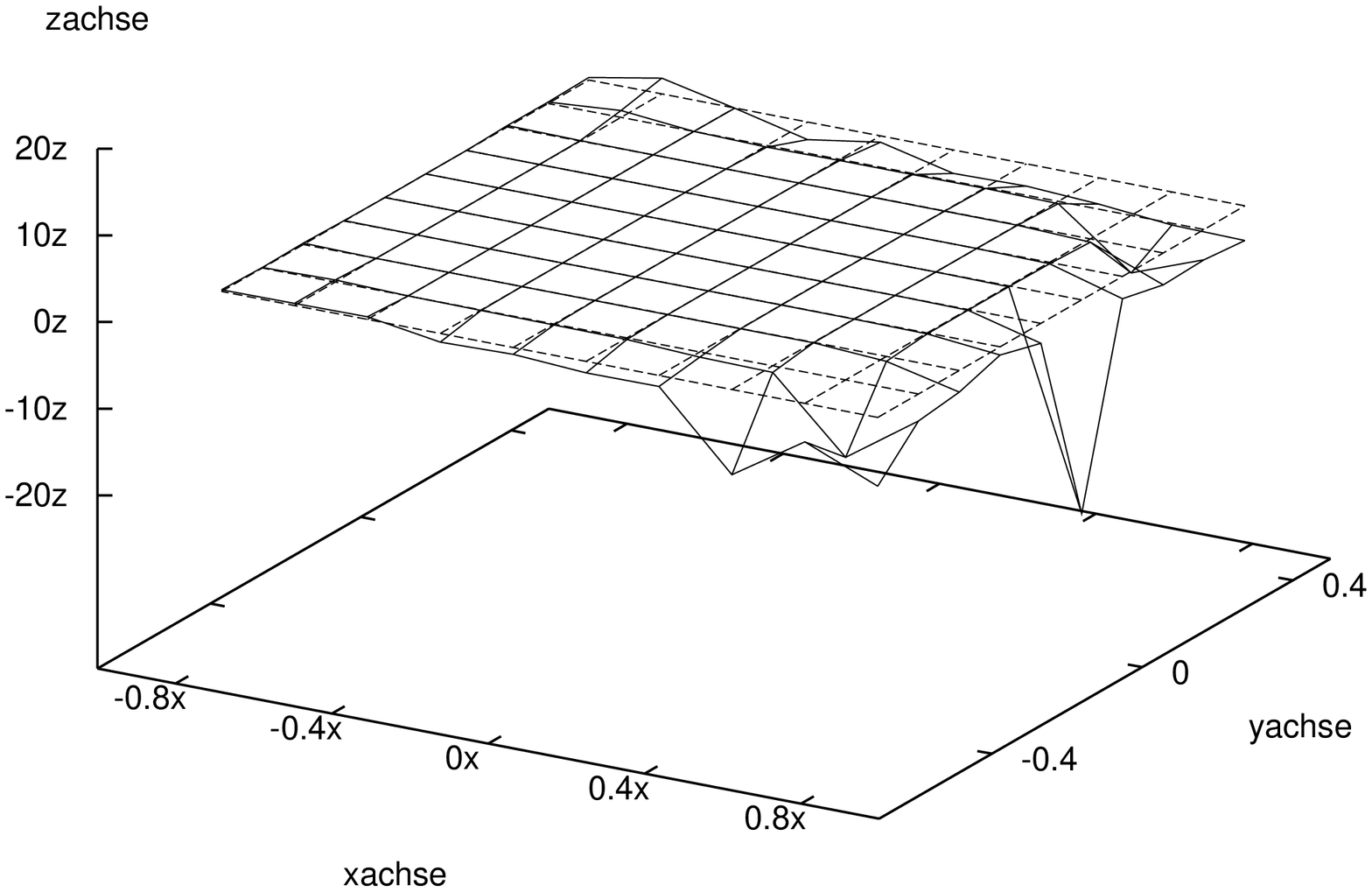}
        \vspace{0.5cm}
      \end{psfrags}
% Unterschrift
      \caption[Diffusion coefficient]{
        {\footnotesize Diffusion coefficient $D^{(2)}_{q_1 q_2}$
          versus variables $q_1$ and $q_2$: As representative for the
          diffusion coefficients, $D^{(2)}_{q_1 q_2}$ has been
          calculated and drawn as lined surface. The dashed
          surface belongs to the expected diffusion coefficient \\
          $D^{(2)}_{q_1 q_2}=0.0025$ }}
      \label{fig:62}
      }
  \end{center}
\end{figure}

Fig. \ref{fig:61} illustrates the drift terms $D_{q_1}^{(1)}({\bf q})$
and $D_{q_2}^{(1)}({\bf q})$, calculated by the presented method,
together with the expected terms for the noisy system
(\ref{II1})-(\ref{II3}). In fig. \ref{fig:62} one of the calculated
diffusion coefficients, $D_{q_1 q_2}^{(2)}({\bf q})$, and its
theoretical function can be seen.

\vspace{1cm}

{\centerline {\bf V. SUMMARY AND OUTLOOK}}
\vspace{0.3cm}

By the discussed method noisy data sets that obey a Fokker-Planck
equation can be analysed and described. By a numerical determination
of the conditional probability distributions drift and diffusion
coefficients can be calculated. Consequently the deterministic laws
and the weight of the fluctuating forces underlying the dynamics of
the system can be extracted.

The algorithm has been illustrated on various one- and
two-di\-men\-sional systems, whose noisy time series have been
simulated and afterwards analysed. A good conformance between
determined and expected drift and diffusion coefficients can be seen.

In biological, physical, technical and economical systems noisy data
sets referring to a Fokker-Planck equation occur very frequently. The
presented method offers the possibility of analysing these systems and
describing them in a mathematical way without the need of any form of
ansatz or assumption. Therefore, there seems to be a broad field of a
possible use of this algorithm.

\end{document}